\RequirePackage{lineno}
\documentclass[twocolumn,showpacs,superscriptaddress,amsmath,amssymb]{revtex4-1}
\usepackage{graphicx}
\usepackage{epsfig}
\usepackage{dcolumn}
\usepackage{bm}
\usepackage{multirow}
\usepackage{overpic}
\usepackage{color}
\usepackage{lineno}
\usepackage{setspace}
\usepackage{mathrsfs}
\usepackage{subfigure}
\include{def-com}
\def \psip {\psi(3686)}
\def \ifb  {\mbox{fb$^{-1}$}}
\def \ipb  {\mbox{pb$^{-1}$}}
\def \jpsi {J/\psi}
\def \epem {e^+e^-}
\def \lplm {\ell^+\ell^-}
\def \mumu {\mu^{+}\mu^{-}}
\def \pppm {\pi^{+}\pi^{-}}
\def \piz {\pi^0}
\def \gev  {\mbox{GeV}}
\def \gevc {\mbox{GeV/$c$}}
\def \gevcc{\mbox{GeV/$c^2$}}
\def \mev  {\mbox{MeV}}
\def \mevcc{\mbox{MeV/$c^2$}}
\def \mfir {(4218.7 \pm 4.0  \pm 2.5)}
\def \wfir {(82.5   \pm 5.9  \pm 0.5)}
\def \msec {(4380.4 \pm 14.2 \pm 1.8)}
\def \wsec {(147.0  \pm 63.0 \pm 25.8)}
\begin{document}
\title{\boldmath Observation of the $Y(4220)$ and $Y(4360)$ in the process $\epem \to \eta\jpsi$}
\author{
\begin{small}
M.~Ablikim$^{1}$, M.~N.~Achasov$^{10,c}$, P.~Adlarson$^{64}$, S. ~Ahmed$^{15}$, M.~Albrecht$^{4}$, A.~Amoroso$^{63A,63C}$, Q.~An$^{60,48}$, ~Anita$^{21}$, Y.~Bai$^{47}$, O.~Bakina$^{29}$
% ...
\\
\vspace{0.2cm}
(BESIII Collaboration)\\
\vspace{0.2cm} {\it
$^{1}$ Institute of High Energy Physics, Beijing 100049, People's Republic of China\\
% ...
$^{4}$ Bochum Ruhr-University, D-44780 Bochum, Germany\\
% ...
$^{10}$ G.I. Budker Institute of Nuclear Physics SB RAS (BINP), Novosibirsk 630090, Russia\\
% ...
$^{15}$ Helmholtz Institute Mainz, Johann-Joachim-Becher-Weg 45, D-55099 Mainz, Germany\\
% ...
$^{21}$ Indian Institute of Technology Madras, Chennai 600036, India\\
% ...
$^{29}$ Joint Institute for Nuclear Research, 141980 Dubna, Moscow region, Russia\\
% ...
$^{47}$ Southeast University, Nanjing 211100, People's Republic of China\\
$^{48}$ State Key Laboratory of Particle Detection and Electronics, Beijing 100049, Hefei 230026, People's Republic of China\\
% ...
$^{60}$ University of Science and Technology of China, Hefei 230026, People's Republic of China\\
% ...
$^{63}$ (A)University of Turin, I-10125, Turin, Italy; (B)University of Eastern Piedmont, I-15121, Alessandria, Italy; (C)INFN, I-10125, Turin, Italy\\
$^{64}$ Uppsala University, Box 516, SE-75120 Uppsala, Sweden\\
% ...
$^{c}$ Also at the Novosibirsk State University, Novosibirsk, 630090, Russia\\
% ...
 }
%\end{center}
\vspace{0.4cm}
%\end{small}
}

\date{\today}
\begin{abstract}
The cross sections of the process $\epem \to \eta\jpsi$ at center-of-mass energies ($\sqrt{s}$) between 3.81 and 4.60~$\gev$ are measured with high precision by using data samples collected with the BESIII detector operating at the BEPCII storage ring.
Three structures are observed by analyzing the lineshape of the measured cross sections, and a maximum-likelihood fit including three resonances is performed by assuming the lowest lying structure is the $\psi(4040)$.
For the other resonances, we obtain masses of $\mfir$ and $\msec$~$\mevcc$ with corresponding widths of  $\wfir$ and $\wsec$~$\mev$, respectively, where the first uncertainties are statistical and the second ones systematic.
The measured resonant parameters are consistent with those of the $Y(4220)$ and $Y(4360)$ from previous measurements of different final states. For the first time, we observe the decays of the $Y(4220)$ and $Y(4360)$ into $\eta\jpsi$ final states.

\end{abstract}

%\pacs{14.40.Rt, 14.40.Pq, 13.66.Bc}

\maketitle
In the past decades, a series of charmonium-like states with $J^{PC}=1^{--}$, so called $Y$ states, were observed in $\epem$ annihilation experiments.
Besides three well-established charmonium states observed in the inclusive hadronic cross section~\cite{PDG}, {\it i.e.}, $\psi(4040)$, $\psi(4160)$, and $\psi(4415)$, five additional states, {\it i.e.}, $Y(4008)$, $Y(4220)$, $Y(4260)$, $Y(4360)$ and $Y(4660)$, were reported in the initial-state radiation (ISR) processes $\epem \to \gamma_{ISR} \pppm\psi$ at the B-factories~\cite{Ystates1,Ystates2,Ystates3,Ystates4,Ystates5,Ystates6,Ystates7,Ystates8} and (or) in the direct production processes $\epem \to \pppm\psi$ at the CLEO and BESIII experiments~\cite{Ystates9,Ystates10},
where the symbol $\psi$ represents both $\jpsi$ and $\psip$ vector charmonium states below the open-charm production threshold.
The latest results of the BESIII experiment for $\epem \to \pppm\jpsi$ show that the $Y(4260)$ may consist of two components, {\it i.e.}, $Y(4220)$ and $Y(4320)$~\cite{Y4220}.
Precise cross section measurements by BESIII in $\epem \to \pppm\psip$ reveal, for the first time, a structure around 4220~$\mevcc$~\cite{pipipsip}.
Dedicated measurements by BESIII of the  cross sections for $\epem\to\pppm h_c$~\cite{pipihc}, $\omega\chi_{c0}$~\cite{omegachic0} and $\pi^+D^0D^{*-}$~\cite{DDpi} also support the existence of structures at 4220 and (or) 4360~$\mevcc$.
Up to now, the internal structure of these $Y$ states are unclear and many theoretical models, such as hybrid charmonium, tetraquark or hadronic molecule, are proposed to interpret their natures, but none of them are conclusive~\cite{Ystates11}.
Complementary measurements of the resonant parameters of the $Y$ states, in particular, by searching for other decay modes will provide further insights into their internal structure.
Hadronic transitions to conventional charmonium states via the emissions of, {\it e.g.}, $\eta$, $\pi^{0}$ or a pion pair, are regarded as sensitive probes~\cite{hadrond}.

The process $\epem \to \eta\jpsi$ was studied using ISR by Belle~\cite{belleetajpsi}. Unlike the process $\epem \to \pppm\jpsi$~\cite{Y4220}, two resonant structures at 4040 and 4160~$\mevcc$, regarded as the $\psi(4040)$ and $\psi(4160)$, respectively, were observed by studying the cross section dependence on the center-of-mass (c.m.)~energy. Furthermore, a hint of an enhancement at 4360~$\mevcc$ was reported.
Using data samples at 17 c.m.~energies from 3.81 to 4.60~$\gev$, BESIII reported more accurate measurements of cross sections of the $\epem \to \eta\jpsi$ process~\cite{etajpsi}.
The BESIII data agree well with that of Belle. However, due to limited statistics it was not possible to establish any potential involved $Y$ states.

In this Letter, we present an updated analysis of $\epem \to \eta\jpsi$ at c.m.~energies between 3.81 and 4.60~$\gev$, where $\jpsi$ is reconstructed with $\lplm$ ($\ell$ = $e$/$\mu$) final states, and $\eta$ is reconstructed via its $\gamma\gamma$ ($\rm mode~I$) and $\pppm\piz$ (mode II) decay modes.
The samples used in this analysis include a set of high luminosity data samples with more than $50~\ipb$ at each c.m.~energy adding up to a total integrated luminosity of 13.1~$\ifb$ (referred to as ``$XYZ$ data'')~\cite{supply}. Compared to the previous analysis~\cite{etajpsi}, 10 high luminosity data sets~\cite{supply}, which are of luminosity greater than $500~\ipb$ individually and with c.m.~energy around 4200~$\mevcc$, are added.
A set of data samples of about 7-9~${\ipb}$ at each c.m. energy with a total integrated luminosity of  0.8~$\ifb$ (named as ``scan data'')~\cite{supply} was used in this study, in addition, which is not available in the earlier study of Ref.~\cite{etajpsi}.
The additional data obtained from mode II, which was absent from the previous measurement~\cite{etajpsi}, are about one quarter of the total statistics collected with mode I.

Details on the features and capabilities of the BEPCII collider and the BESIII detector can be found in Ref.~\cite{besint}. The {\sc geant4}-based~\cite{geant4} Monte Carlo (MC) simulation software package {\sc{boost}}~\cite{boostc},
which includes the geometric description of the BESIII detector and the detector response,
is used to optimize event selection criteria, determine the detection efficiencies,
and estimate the background events.
Signal MC samples of $\epem\to\eta\jpsi$ with the corresponding $\jpsi$ and $\eta$ decay modes %$\eta\to\gamma\gamma$/$\pi^{+}\pi^{-}\pi^{0}$, $\jpsi\to\lplm$ ($\ell$ = $e$/$\mu$)
are generated using {\sc{helamp}}~\cite{ampli} and {\sc{evtgen}}~\cite{Evtgen} at each c.m.~energies.
% the helicity amplitude model where the angular correlation between $\eta$ decay and $\jpsi$ has been considered.
The ISR is simulated with {\sc{kkmc}}~\cite{kkmc} by requiring a maximum energy of the ISR photon corresponding to the $\eta\jpsi$ mass threshold.
Final-state radiation (FSR) is simulated with {\sc{photos}}~\cite{photon}.
Possible background contributions are studied with the inclusive MC samples generated by {\sc{kkmc}} with comparable luminosity to the $XYZ$ data, where the known decay modes are simulated by {\sc{evtgen}}~\cite{Evtgen} with branching fractions taken from the PDG~\cite{PDG}, and the remaining unknown are simulated with the {\sc{lundcharm}} model~\cite{lumarlw}.

Candidate events are required to have two (with zero net charge) charged tracks for mode I and four (with zero net charge) charged tracks for mode II.
The charged tracks are reconstructed with the hit information in the main drift chamber (MDC).
Each charged track is required to be within the polar angle ($\theta$) ranging $|\cos\theta|<0.93$
and to have a point of closest approach to the interaction point (IP) within $\pm10$~cm
along the beam direction and 1~cm in the plane perpendicular to the beam.
Since pions and leptons have distinct momenta for the signal processes, we assigned charged tracks with momenta ($p$) larger than 1.0~$\gevc$ to be leptons, otherwise as pions.
The separation of electrons from muons is realized by taking into account the deposited energy ($E$) in the electromagnetic calorimeter (EMC), {\it i.e.}, a muon is required to have $E<0.5~\gev$, while electron to be $E/p>0.8$.
The signal candidates are required to have a pair of lepton with same flavor but opposite charge.
In mode II, two additional pions with opposite charge are further required.
Photon candidates are reconstructed by isolated clusters in the EMC, which are at least 10$^{\circ}$ away from the nearest charged track.
The photon energies are required to be at least 25~$\mev$ in barrel region ($|\cos\theta|$~$<$~0.8), or 50~$\mev$ in end-cap regions (0.86~$<$~$|\cos\theta|$~$<$~0.92).
A requirement on the EMC timing (0~$<t<$~700 ns) is implemented to suppress electronic noise and energy deposition unrelated to the event.
Candidate events with at least two photons are taken for further analysis.

To improve the kinematic resolution and to suppress the background events, a four-constraint (4C) kinematic fit imposing energy-momentum conservation with the hypothesis of $\epem\to \gamma\gamma\lplm$ is applied for the candidates of mode I, while a five-constraint (5C) kinematic fit is performed under the hypothesis of $\epem\to \gamma\gamma\pppm\lplm$ with additional $\piz$ mass constraint for the photon pair of the mode II.
For events with more than two photon candidates, all photon pairs are tested in the kinematic fit and the combination with the smallest $\chi^{2}_{\rm{4C/5C}}$ is retained.
The surviving events are further required to satisfy $\chi^{2}_{\rm 4C} <$ 40 or $\chi^{2}_{\rm 5C} <$ 80. To further suppress the background events from the radiative Bhabha and dimuon events associated with a random photon candidate for events of the mode I, the energy of each of two selected photons is required to be larger than 80~$\mev$.

Figure~\ref{fig:2d} presents the distributions of the invariant mass of the $\lplm$ pair ($M(\ell^+\ell^-)$) versus that of the $\gamma\gamma$ pair ($M(\gamma\gamma)$) or $\pppm\piz$ combination ($M(\pppm\piz)$) for the surviving events at $\sqrt{s}= 4.1780~\gev$ after applying the previously described selection criteria.
Clear accumulations of candidate events of the signal process $\epem\to\eta\jpsi$ are observed around the intersections of the $\jpsi$ and $\eta$ mass regions.
Signal candidates are required to be within the $\jpsi$ mass region, defined as [3.067, 3.127]~$\gevcc$, which is approximately 3$\sigma$ of the resolution of the invariant mass distributions of lepton pairs.
The events in the $\jpsi$ mass sideband regions, defined as [3.027, 3.057] and [3.137, 3.167]~$\gevcc$, are used to estimate the non-$\jpsi$ background, and non-peaking background are observed in the $M(\gamma\gamma)$ and $M(\pppm\piz)$ distributions.
A significantly larger non-$\jpsi$ background is observed in the $\epem$ mode than in the $\mumu$ mode in mode I, which is due to the large Bhabha cross section.

  \begin{figure}[htbp]
 \centering
 \begin{overpic}[width=0.22\textwidth]{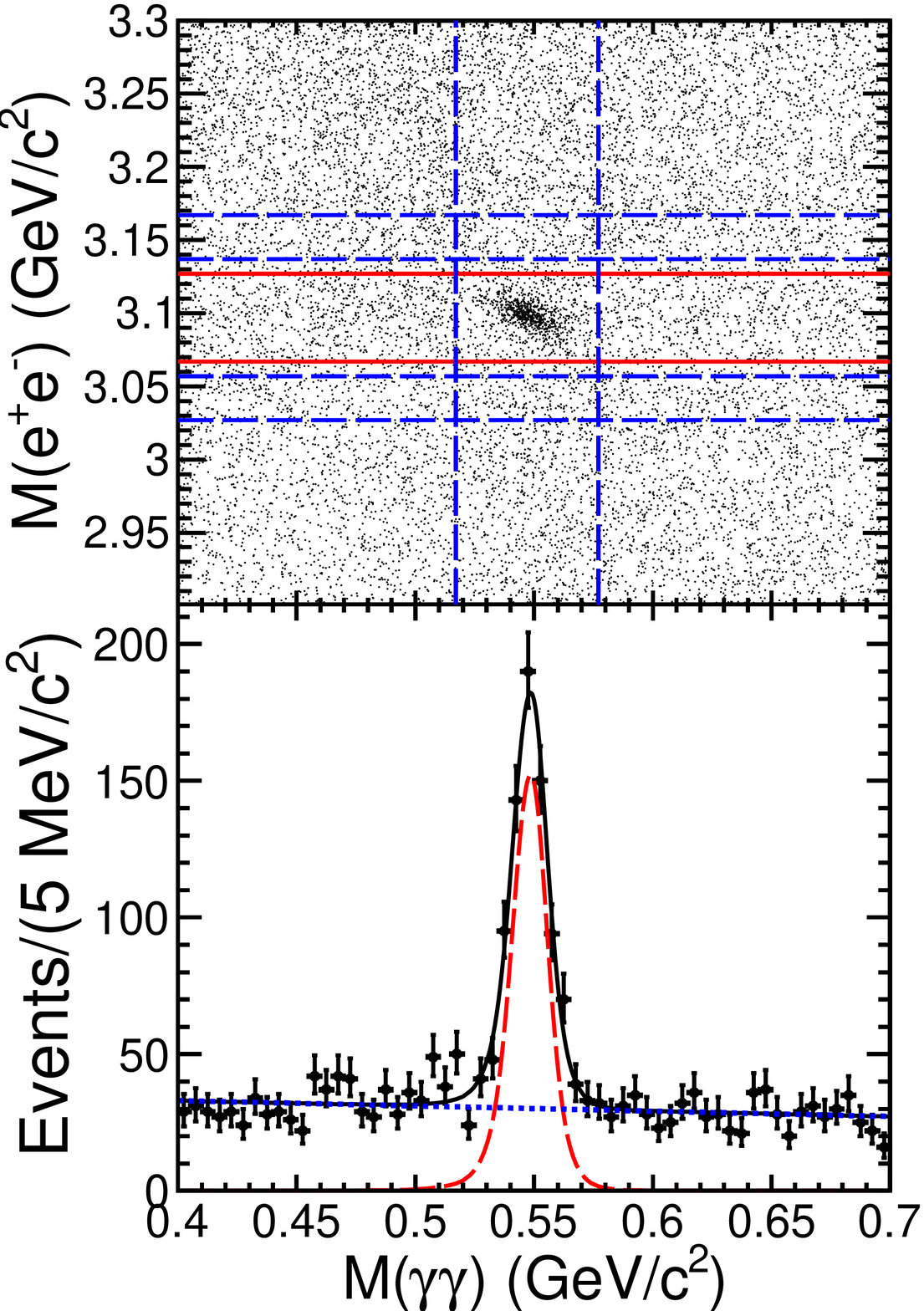}
 \put(55,90){ (a)}
 \put(55,40){ (c)}
 \end{overpic}
 \begin{overpic}[width=0.22\textwidth]{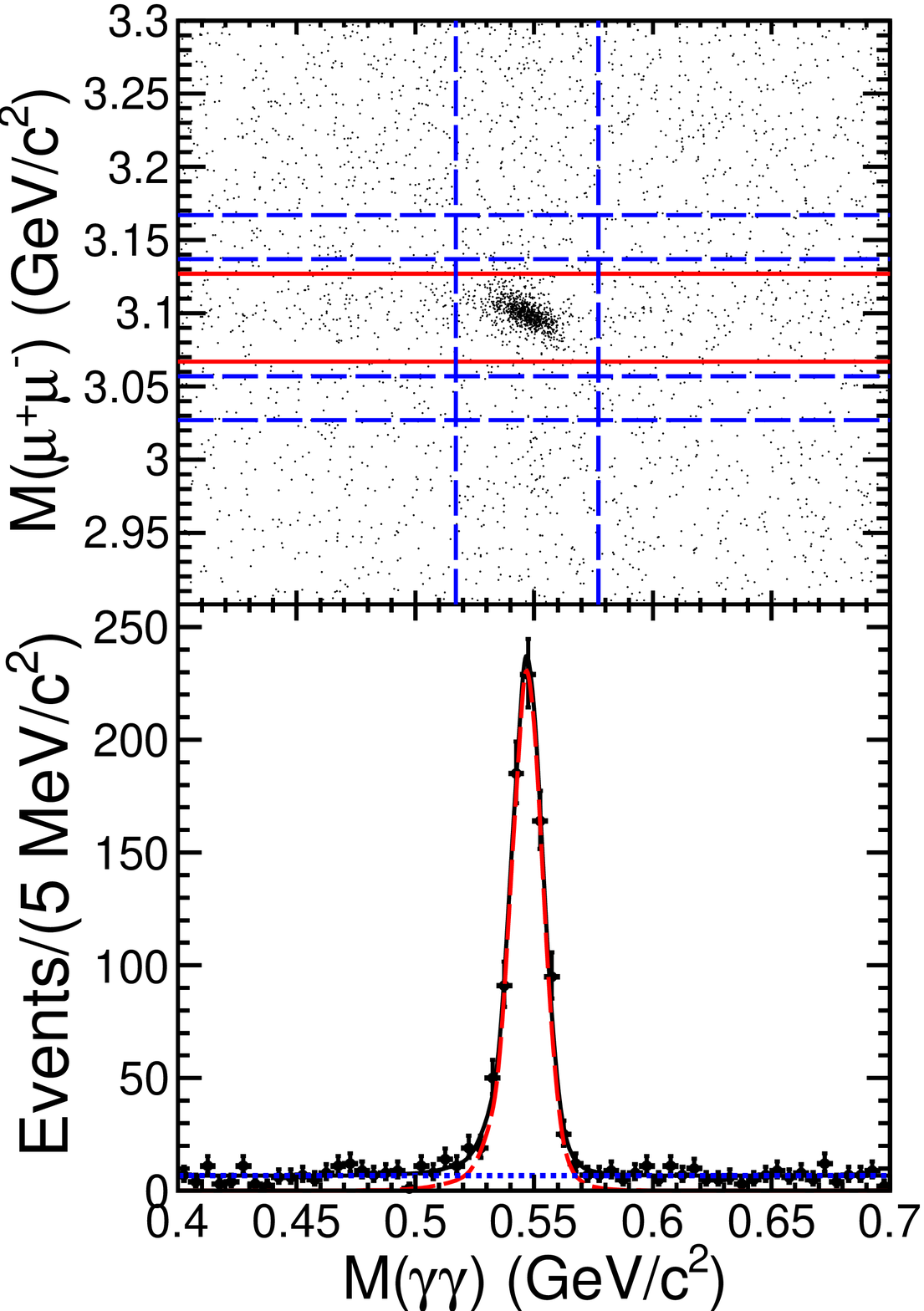}
 \put(55,90){ (b)}
 \put(55,40){ (d)}
 \end{overpic}
 \begin{overpic}[width=0.22\textwidth]{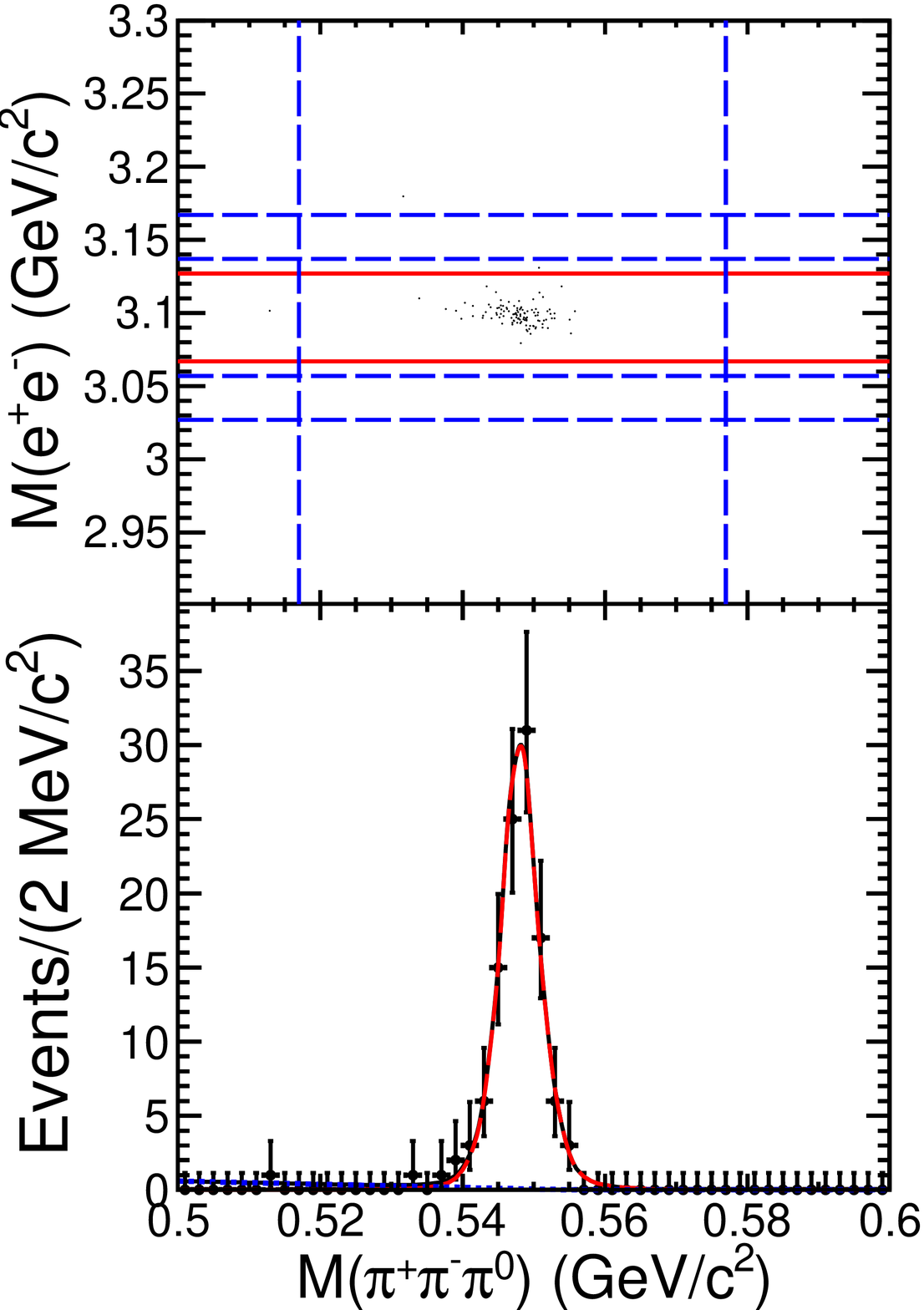}
 \put(55,90){ (e)}
 \put(55,40){ (g)}
 \end{overpic}
  \begin{overpic}[width=0.22\textwidth]{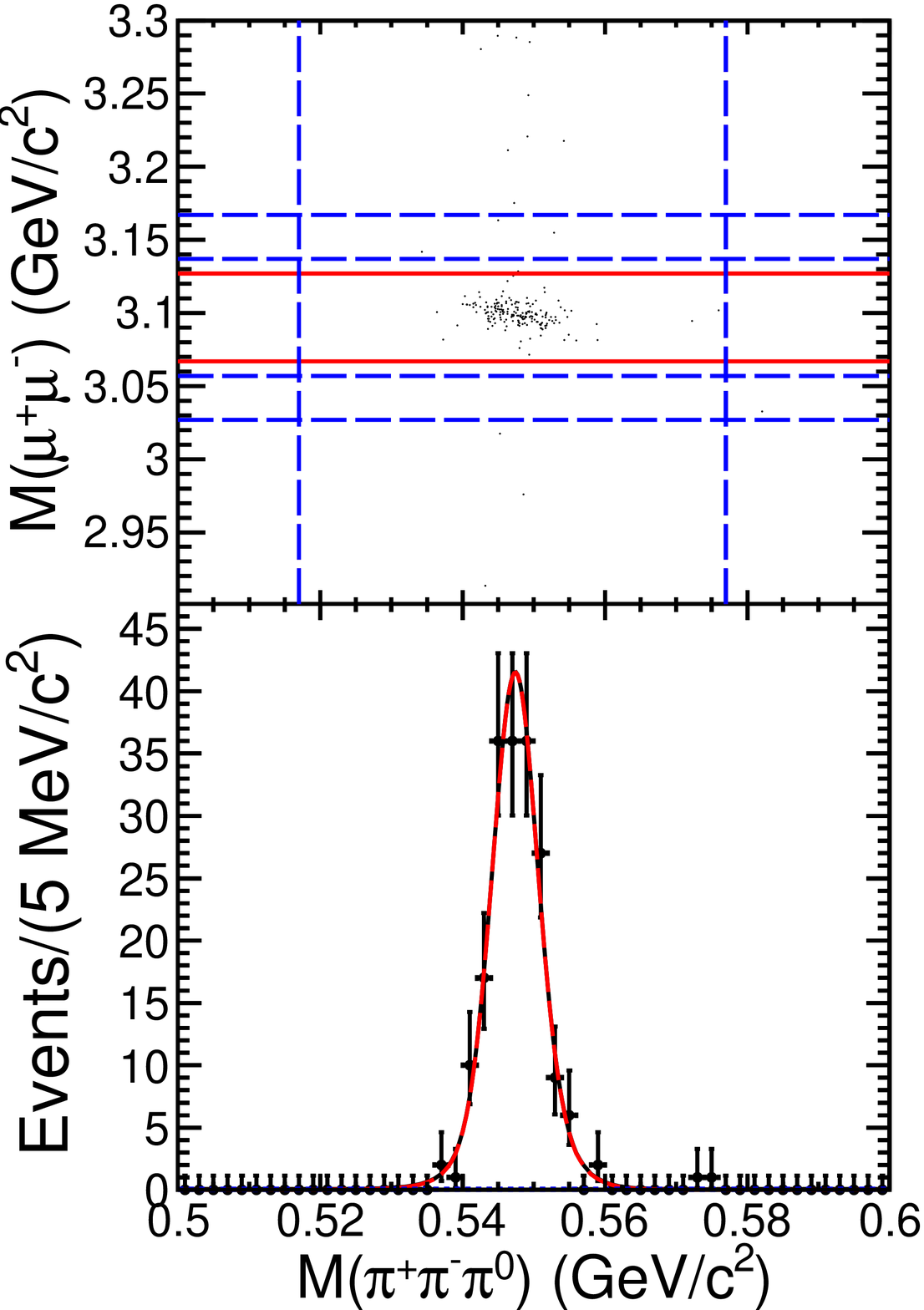}
 \put(55,90){ (f)}
 \put(55,40){ (h)}
 \end{overpic}
 \caption{ (Color online) Scatter plots of $M$($\lplm$) versus $M$($\gamma\gamma/\pppm\piz$) (a, b, e, f) and the spectra of the $M$($\gamma\gamma/\pppm\piz$) distribution (c, d, g, h) in the $\jpsi$ signal region for data at $\sqrt{s}$ = 4.1780~$\gev$. The upper panels correspond to the mode I and those at bottom for the mode II.
 In the scatter plots, the solid (red) lines denote the signal region, and the dashed (blue) lines for the sideband region.
 For the mass spectra plots, the dots with error bars represent data. The solid curves correspond to the fit results whereby the long dashed (red) curves for the signal and the short dashed (blue) curves for background.
}
  \label{fig:2d}
 \end{figure}

The Born cross section is obtained from
   \begin{equation}
     \sigma^{B}=\frac{N^{sig}}{\mathcal{L}_{int}\cdot(1+\delta)^{r}\cdot(1+\delta)^{v}\cdot\mathcal{B}r\cdot\epsilon},
     \label{equation}
   \end{equation}
where $N^{sig}$ is the signal yield, which will be explained below,
$\mathcal{L}_{int}$ is the integrated luminosity,
$(1+\delta)^{r}$ is the ISR correction factor,
$(1+\delta)^{v}$ is the vacuum polarization factor taken from a QED calculation~\cite{VP},
$\mathcal{B}r$ is the product of the branching fractions of the subsequent decays of intermediate states quoted from the PDG~\cite{PDG},
and $\epsilon$ is the detection efficiency obtained from a MC simulation.
The ISR correction factor is obtained by using iteratively the QED calculation as described in Ref.~\cite{VR},
where the last measured cross section is taken as the input lineshape.

For the $XYZ$ data, an unbinned maximum-likelihood fit is performed on the distributions of $M(\gamma\gamma)$ and $M(\pppm\piz)$ to extract the signal yields, where the signal is described by a MC-simulated shape convolved with a Gaussian function, representing the resolution difference between data and MC simulation, and the background is described by a linear function.
A simultaneous fit is performed by considering the four processes, ${\it i.e.}$ two observation variables $M(\gamma\gamma)$ and $M(\pppm\piz)$, as well as two $\jpsi$ decay modes $\epem$ and $\mumu$ for the 27 data samples at different c.m.~energies.
In the fit, the different processes are constrained by the same Born cross section $\sigma^B$, and  the expected signal yields are
$N^{sig}=\sigma^{B}\cdot\mathcal{L}_{int}\cdot(1+\delta)^{r}\cdot(1+\delta)^{v}\cdot\mathcal{B}r\cdot\epsilon$.
Among the different data sets we used common fit parameters for the mean and width of the Gaussian function representing differences between data and MC.
For center of mass energies where the signal is not significant, we compute the upper limits on the cross sections at 90\% C.L. using a Bayesian method with a flat prior. The optimized likelihoods $\mathscr{L}$ are presented as a function of the cross section. The upper limits on the cross section $\sigma^{UP}$ at 90\% C.L. are the values that yield 90\% of the likelihood integral over $\sigma$ from zero to infinity: $\int^{\sigma^{UP}}_{0}\mathscr{L}d\sigma$/$\int^{\infty}_{0}\mathscr{L}d\sigma$=0.9. The systematic uncertainty is taken into account by smearing the posterior distribution.

For the scan data sets, the signal yields are determined by counting the number of events in the signal region after subtracting the background estimated by the normalized number of events in the $\jpsi$ mass sideband region. Only the mode I is considered to extract the Born cross sections by using Eq.~(\ref{equation}).

The measured Born cross sections at the different c.m.~energies for both $XYZ$ and scan data are shown in
the top and bottom panels of Fig.~\ref{fit_etajpsixyz}, respectively. Clear structures are observed.
The numbers used in the calculation of the Born cross section (upper limit at 90\% C.L.) are summarized in Tables I and II in the supplemental material~\cite{supply}.

\begin{figure}[htbp]
  \centering
  % Requires \usepackage{graphicx}

  \begin{overpic}[width=0.4\textwidth]{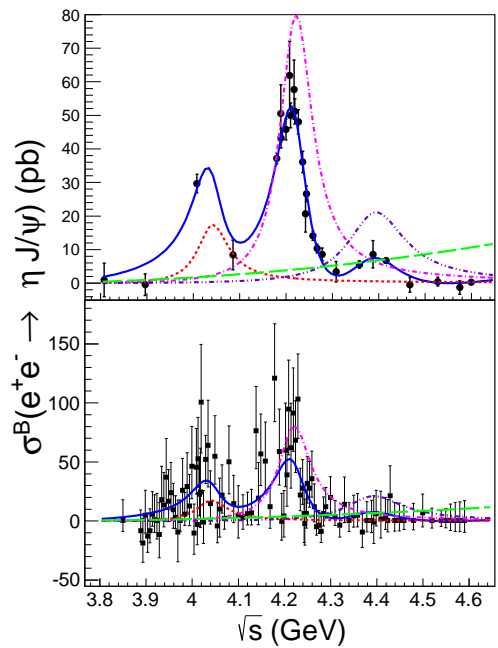}
%  \put(1,50){ \rotatebox{90}{$\sigma(e^{+}e^{-} \rightarrow \eta \jpsi) (pb)$}}
  \end{overpic}

   \caption{(Color online) Fit to the cross section data of $\epem\to \eta \jpsi$ for $XYZ$ data (upper) and scan data (bottom).
   Dots with error bars are data.
   The solid curves (blue) are the fit results;
   the dashed curves (red) for $\psi(4040)$; the short-dashed curves (pink) for $Y(4220)$; the double short-dashed curves (purple) for $Y(4360)$; the long-dashed curve (green) for the P-PHSP component.
   %For $XYZ$ data, the short and long error bars are the statistical and total (combined statistical and systematic) uncertainties, respectively.
   }\label{fit_etajpsixyz}

\end{figure}

The following sources of systematic uncertainty are considered in the cross section measurements.
%The uncertainties of detection efficiency include 1\% for each charged track~\cite{trackerror} and photon~\cite{photonerror}, individually.
The uncertainty of the integrated luminosity is 1\% measured by analyzing events of the Bhabha scattering process~\cite{lum}.
The uncertainty related with the efficiencies of leptons, pions and photons is 1\% for each particle~\cite{trackerror,photonerror}.
%The uncertainty associated with the efficiency of pion pair (presented in the mode II only) is assigned to be 1\% each tracks~\cite{trackerror}.
The uncertainties related to the $\jpsi$ mass window requirement and kinematic fit are
estimated by tuning the MC sample for the $\jpsi$ mass resolution and the helix parameters of charged tracks~\cite{helixsys} according to data,
%The uncertainty associated with the kinematic fit is estimated by correcting the helix parameters of the charged tracks of MC sample to match its resolutions~\cite{helixsys}.
and taking the resulting changes in efficiency as the uncertainties.
The uncertainty associated with ISR correction factor is taken to be the difference of $(1+\delta)^{r}\cdot\epsilon$ between the last two iterations in the cross section measurement.
The uncertainties of the branching fractions of the intermediate states are taken from the PDG~\cite{PDG}.
As described above, the signal yields are extracted by performing a simultaneous fit, thus, those uncertainties, which are correlated ({\it i.e.} luminosity, lepton and photon efficiencies), are directly propagated to the measured cross sections.
Otherwise, we repeated the simultaneous fits by changing the corresponding value by $\pm 1\sigma$, individually, and the largest changes in the results are taken as the uncertainties.
To extract the uncertainties associated with the fit procedure, we perform alternative fits by replacing the linear function with a second-order polynomial function for the background, fixing the width of the Gaussian function for the signal to be its nominal value and in addition changing its uncertainty and varying the fit range, individually, and the relative changes in the results are taken as the uncertainties.
The efficiencies for the other selection criteria, the trigger simulation, the event start time determination, and the FSR simulation, are quite high ($>$99\%), and their systematic errors are estimated to be less than 1\%.
Assuming all sources of uncertainties are independent, the total uncertainties in the $\eta\jpsi$ cross section measurement are determined to be 3.5 - 13.7\% depending on the c.m. energy. In general, the systematical errors are much smaller than the statistical ones. For details, we refer to Table III of the supplemental material~\cite{supply}.

To extract the resonant parameters of the structures observed in the measured cross sections, a simultaneous maximum-likelihood fit is performed to the results extracted from the $XYZ$ and scan data.
The fit function is a coherent sum of a P-wave phase space component (P-PHSP) ($\Phi(\sqrt{s})$) of the process  $\epem\to\eta\jpsi$  and three Breit-Wigner functions ($B_{i=1,2,3}$)
for the structures observed around 4040, 4230 and 4360~$\mevcc$, respectively:
 \begin{equation}
 \begin{split}
\sigma^{B}(\sqrt{s})= |C_{0}\sqrt{\Phi(\sqrt{s})} + e^{i\phi_1}B_{1}(\sqrt{s}) + \\
                   e^{i\phi_2}B_{2}(\sqrt{s}) + e^{i\phi_3}B_{3}(\sqrt{s})|^{2},
 \label{eq2}
 \end{split}
\end{equation}
where $\phi_i$ is the relative phase of a resonance ($i$) to the P-PHSP component, $C_{0}$ is a free parameter.
The parameterizations of the P-PHSP and Breit-Wigner components are formualted as
\begin{equation}
\Phi(\sqrt{s})=\frac{q^{3}}{\sqrt{s}},
\label{eq3}
\end{equation}
 \begin{equation}
       B_{i}(\sqrt{s}) = \frac{M_i}{\sqrt{s}}\frac{\sqrt{12\pi {\mathcal{B}r}_i \Gamma^{e^{+}e^{-}}_i \Gamma_i}}{s-M_i^{2}+iM_ i\Gamma_i}\sqrt{\frac{\Phi(\sqrt{s})}{\Phi(M_i)}},
 \label{eq4}
 \end{equation}
where $q$, $M_i$, $\Gamma_i$, $\Gamma_{i}^{\epem}$ and $\mathcal{B}r_i$ are the daughter momentum in the rest frame of its parent, the mass, width, partial width of the decay in $\epem$, and the branching fraction to $\eta\jpsi$ mode for the resonance $i$, respectively.

The fit is carried out by incorporating the statistics uncertainties only, where the number of events for the scan data are assumed to be Poisson distribution, and those for the XYZ data are Gaussian distribution.
Additionally, the beam energy spread of BEPCII (1.6 MeV) is considered by convolving with a Gaussian function whose width is 1.6 MeV~\cite{pipihc,beamspread}.
%the beam energy spread of BEPCII (1.6~$\mev$)~\cite{pipihc,beamspread} is considered in the fit by weighting the expected cross section with a Gaussian distribution of beam energy. 
%The fit is carried out by considering the spread of beam energy (1.6~$\mev$). where the number of events in signal and sideband regions for the scan data is assumed to be Poisson distribution, and others for the XYZ data follow Gaussian distributions~\cite{pipihc}.
The structure around 4040~$\mevcc$ is assumed to be the $\psi(4040)$, and its mass and width are fixed to those given in the PDG~\cite{PDG}, due to a lack of data sets at this energy region. %its mass and width are fixed to those given in the PDG~\cite{PDG}.
Three solutions are found with equal fit quality and with identical masses and widths for the structures around 4220 and 4360~$\mevcc$. The fit quality is $\chi^{2}$/n.d.f. = 107.7/120, estimated by a $\chi^{2}$-test approach, %and resulted in $\chi^{2}$/n.d.f. = 107.7/120,
where n.d.f. is the number of degree of freedom.
The fit results are summarized in Table~\ref{results3} and the fit curves (Solution 1) are exhibited in Fig.~\ref{fit_etajpsixyz}.

%To estimate the significance of the structure, the fit is performed by considering the correlated and uncorrelated uncertainties of measured Born cross section, where the correlated uncertainties are considered by convolving with a Gaussian function whose mean and width is 0 and 3.3 respectively, the other uncertainties are considered by adding these uncertainties into the statistical ones in quadrature.
%The statistical significance of the structure around 4360~$\mevcc$ is 6.0$\sigma$, which is obtained by performing an alternative fit without this structure included, and calculated by incorporating the ratio of the logarithmic likelihood value which follows $\chi^{2}$ distribution and of n.d.f. relative to the nominal fit.}
%The significance of the structures, which is obtained by performing an alternative fit without this structure included, and calculated by incorporating the ratio of the logarithmic likelihood value which follows $\chi^{2}$ distribution and of n.d.f. relative to the nominal fit. The significance of the structures around 4220~$\mevcc$ and around 4360~$\mevcc$ is estimated to be 9.1 $\sigma$ and 6.0$\sigma$, respectively.

\begin{table}[htbp]
\centering
\footnotesize
\caption{ Fitting results of the  $\epem \to \eta \jpsi$ decay.}
\label{results3}
    \begin{tabular}{l | ccc}
      \hline
  \hline
    Parameters                 &~~Solution 1~~    &~~Solution 2~~     &~~Solution 3~~   \\
  \hline
$ M_{1}(\mevcc)$              &\multicolumn{3}{c}{4039(fixed)}                \\
$\Gamma_{1}(\mev)$          &\multicolumn{3}{c}{80(fixed)}                  \\
$\Gamma_{1}^{\epem}\mathcal{B}r_1$(eV)  & $1.6\pm0.3$ 	           & $1.5\pm0.3$      	& $7.1\pm0.6$ 	 \\
$\phi_{1}$(rad)             & $3.3\pm0.3$ 		       & $3.1\pm0.3$        & $4.5\pm0.2$     \\
\hline
$ M_{2}(\mevcc)$              &  \multicolumn{3}{c}{4218.7 $\pm$4.0}  \\
$\Gamma_{2}(\mev)$          &  \multicolumn{3}{c}{82.5 $\pm$ 5.9}     \\
$\Gamma_{2}^{\epem}\mathcal{B}r_2$(eV)  & $8.2\pm1.8$   	& $4.9\pm1.0$               & $7.2\pm1.5$  \\
$\phi_{2}$(rad)             & $4.3\pm0.4$ 		& $3.6\pm0.3$               & $3.0\pm0.3$ \\
\hline
$M_{3}(\mevcc)$               & \multicolumn{3}{c}{4380.4 $\pm$ 14.2 }      \\
$\Gamma_{3}(\mev)$          & \multicolumn{3}{c}{147.0 $\pm$ 63.0}         \\
$\Gamma_{3}^{\epem}\mathcal{B}r_3$(eV)  & $3.9\pm2.6$ 	   & $1.7\pm1.1$                & $2.0\pm1.4$  \\
$\phi_{3}$(rad)             & $2.8\pm0.4$        & $3.3\pm0.4$             & $3.0\pm0.4$  \\
  \hline
  \hline
    \end{tabular}
\end{table}

%In the following, we discuss the main sources of
The systematic uncertainties of the resonant parameters of the structures at 4220 and 4360~$\mevcc$ and of the product of $\mathcal{B}r_i$ and $\Gamma^{e^{+}e^{-}}_i$ are discussed as follows.
The uncertainties associated with the systematic of measured cross section are estimated by incorporating the correlated and uncorrelated systematic uncertainties of measured Born cross section in the fit.
The uncertainty associated with the c.m.~energy (0.8~$\mev$)~\cite{Y4220} is common for all data samples and propagates directly to the mass measurement.
The uncertainty associated with the fit range is investigated by excluding the last energy point $\sqrt{s} =4.60~\gev$ in the fit.
The uncertainty from the $\psi(4040)$ resonant parameters is studied by varying the parameters within its uncertainties.
We performed the alternative fits with above scenarios, individually, the resultant difference are taken as the systematic uncertainties, and are summarized in Table~\ref{sys}.

The structure at 4360~$\mevcc$ is firstly observed in the process $\epem\to\eta \jpsi$, the corresponding significance is studied by performing an alternative fit without this structure included. 
The significance is 6.0~$\sigma$, calculated with the change of likelihood values and of n.d.f. relative to the nominal fit and incorporated all the uncertainties discussed above.

%To estimate the significance of the structure, the fit is performed by considering the correlated and uncorrelated n.d.f. relative to the nominal fit. uncertainties of measured Born cross section, where the correlated uncertainties are considered by convolving with a Gaussian function whose mean and width is 0 and 3.3 respectively, the other uncertainties are considered by adding these uncertainties into the statistical ones in quadrature.
%The statistical significance of the structure around 4360~$\mevcc$ is 6.0$\sigma$, which is obtained by performing an alternative fit without this structure included, and calculated by incorporating the ratio of the logarithmic likelihood value which follows $\chi^{2}$ distribution and of n.d.f. relative to the nominal fit.}
%The significance of the structures, which is obtained by performing an alternative fit without this structure included, and calculated by incorporating the ratio of the logarithmic likelihood value which follows $\chi^{2}$ distribution and of n.d.f. relative to the nominal fit. The significance of the structures around 4220~$\mevcc$ and around 4360~$\mevcc$ is estimated to be 9.1 $\sigma$ and 6.0$\sigma$, respectively.

\begin{table}[htbp]
\centering

\caption{ Systematic uncertainties of the resonant parameters of structure around 4220 and 4360~$\mevcc$ and of the product of $\mathcal{B}r_i$ and $\Gamma^{e^{+}e^{-}}_i$.}
\resizebox{8.5cm}{2.2cm}{
    \begin{tabular}{l |c ccccc}
      \hline
  \hline
Sources              & Solution & $\sqrt{s}$   &~Fit range~ &~$\psi(4040)$~    & Cross section &  Total    \\ \hline
$M_{1}(\mevcc)$     & -- & 0.8          & 0.1        &  0.2                  & 2.4  &  2.5      \\
$\Gamma_{1}(\mev)$& -- &  --            & 0.1        &  0.4                  & 0.3  &  0.5      \\
$M_{2}(\mevcc)$     & -- & 0.8          & 0.3        &  1.3                  & 0.9  &  1.8      \\
$\Gamma_{2}(\mev)$& -- & --             & 0.6        &  3.0                  & 25.6 &  25.8      \\

\hline
 \multirow{3}{*}{$\Gamma_{1}^{\epem}\mathcal{B}r_1$(eV)}  &Solution 1& -- &  0.1  & 0.1 & 0.1 &   0.2 \\
                                                 &Solution 2& -- &  0.1  & 0.1          & 0.1 &   0.2 \\
                                                 &Solution 3& -- &  0.1  & 0.5          & 0.2 &   0.5 \\
\hline
 \multirow{3}{*}{$\Gamma_{2}^{\epem}\mathcal{B}r_2$(eV)}  &Solution 1& -- &  0.1  & 0.1 & 0.6 &   0.6 \\
                                                 &Solution 2& -- &  0.1  & 0.1          & 0.1 &   0.2 \\
                                                 &Solution 3& -- &  0.1  & 0.4          & 0.1 &   0.4 \\
\hline
 \multirow{3}{*}{$\Gamma_{3}^{\epem}\mathcal{B}r_3$(eV)}  &Solution 1& -- &  0.1  & 0.2 & 0.9 &   0.9 \\
                                                 &Solution 2& -- &  0.1  & 0.1          & 0.5 &   0.5 \\
                                                 &Solution 3& -- &  0.1  & 0.5          & 0.6 &   0.8 \\
  \hline
  \hline
    \end{tabular}
}
    \label{sys}
\end{table}

In summary, we measured the Born cross sections of $\epem \to \eta\jpsi$ for c.m.~energy between 3.81 and 4.60~GeV by using the data samples collected by the BESIII experiment.
The measured cross sections are fitted by including three resonant structures and assuming the lowest lying one is the $\psi(4040)$.
The masses and widths of the two resonances are found to be $\mfir$ and $\msec$~$\mevcc$ and the width $\wfir$ and $\wsec$~$\mev$, respectively,
where first uncertainties are statistical and second ones systematic.
It should be noted that we found a resonant structure with a mass around 4220~$\mevcc$ that is significantly higher than the one (4160~$\mevcc$) observed by the Belle experiment~\cite{belleetajpsi}.
A comparison of masses versus widths for the structures in this measurement as well as those obtained from the processes $\epem\to\pppm\jpsi$~\cite{Y4220}, $\pppm\psip$~\cite{pipipsip}, $\pppm h_c$~\cite{pipihc}, $\omega\chi_{c0}$\cite{omegachic0} and $\pi^{+}D^{0}D^{*-}$~\cite{DDpi} by the BESIII experiment are presented in Fig.~\ref{fig:3}.
The measured resonant parameters of the two observed structures are consistent with or close to those of previous measurements, however, the intrinsic scenario for the difference on width is still unknown.
Assuming that the two observed structures are the $Y(4220)$ and $Y(4360)$,
%this is for the first time we observe these two $Y$ states decaying into $\eta\jpsi$ final states.
our result would be the first mass, width, and branching fraction measurements of the two $Y$ states decaying into the $\eta\jpsi$ final state.

\begin{figure}[htbp]
  \centering
  % Requires \usepackage{graphicx}

 % \includegraphics[width=0.35\textwidth]{fig/Y423068.eps}
  \includegraphics[width=0.4\textwidth]{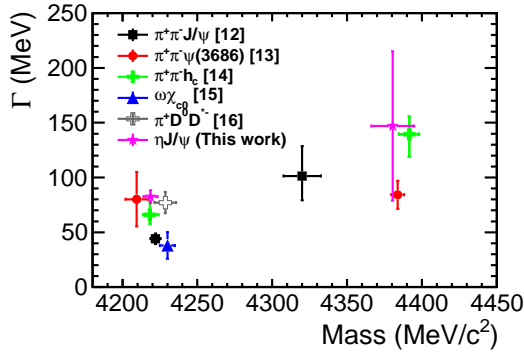}

   \caption{ Masses versus widths of the $Y(4220)$ and $Y(4360)$ obtained from the different final states by BESIII.}\label{fig:3}
\end{figure}

The BESIII collaboration thanks the staff of BEPCII and the IHEP computing center and the supercomputing center of USTC for their strong support. This work is supported in part by National Key Basic Research Program of China under Contract No. 2015CB856700; National Natural Science Foundation of China (NSFC) under Contracts Nos. 11625523, 11635010, 11735014, 11822506, 11835012, 11935015, 11935016, 11935018, 11961141012, 11335008, 11375170, 11475164, 11475169, 11625523, 11605196, 11605198, 11705192; the Chinese Academy of Sciences (CAS) Large-Scale Scientific Facility Program; Joint Large-Scale Scientific Facility Funds of the NSFC and CAS under Contracts Nos. U1732263, U1832207, U1532102, U1732263, U1832103; CAS Key Research Program of Frontier Sciences under Contracts Nos. QYZDJ-SSW-SLH003, QYZDJ-SSW-SLH040; 100 Talents Program of CAS; INPAC and Shanghai Key Laboratory for Particle Physics and Cosmology; ERC under Contract No. 758462; German Research Foundation DFG under Contracts Nos. Collaborative Research Center CRC 1044, FOR 2359; Istituto Nazionale di Fisica Nucleare, Italy; Ministry of Development of Turkey under Contract No. DPT2006K-120470; National Science and Technology fund; STFC (United Kingdom); The Knut and Alice Wallenberg Foundation (Sweden) under Contract No. 2016.0157; The Royal Society, UK under Contracts Nos. DH140054, DH160214; The Swedish Research Council; U. S. Department of Energy under Contracts Nos. DE-FG02-05ER41374, DE-SC-0012069
%\begin{thebibliography}{99}

\end{document}